\documentclass[journal=jacsat,manuscript=article]{achemso}

\usepackage[version=3]{mhchem} 
\usepackage{amsfonts}
\usepackage{wrapfig}
\usepackage{placeins}
\usepackage{xcolor}

\author{Aashwin Mishra}
\affiliation{Machine Learning Initiative, SLAC National Accelerator Laboratory, USA}
\author{Daniel Ratner}
\affiliation{Machine Learning Initiative, SLAC National Accelerator Laboratory, USA}
\author{Quynh L. Nguyen}
\affiliation{Linac Coherent Light Source, SLAC National Accelerator Laboratory, USA}
\alsoaffiliation{Fundamental Physics Directorate, SLAC National Accelerator Laboratory, USA}
\alsoaffiliation{Current address: NVIDIA, Santa Clara, USA}
\email{qlnguyen@slac.stanford.edu}

\title[An \textsf{achemso} demo]
  {Deep Learning Enabled Nanoscale X-ray Photoemission Electron Microscopy (nanoXPEEM)}

\abbreviations{X-ray photoemission, microscopy, Deep Learning, 2D materials}

\begin{document}

\begin{tocentry}

Some journals require a graphical entry for the Table of Contents.
This should be laid out ``print ready'' so that the sizing of the
text is correct.

Inside the \texttt{tocentry} environment, the font used is Helvetica
8\,pt, as required by \emph{Journal of the American Chemical
Society}.

The surrounding frame is 9\,cm by 3.5\,cm, which is the maximum
permitted for  \emph{Journal of the American Chemical Society}
graphical table of content entries. The box will not resize if the
content is too big: instead it will overflow the edge of the box.

This box and the associated title will always be printed on a
separate page at the end of the document.

\end{tocentry}

\begin{abstract}

Understanding and manipulating two-dimensional materials for real-world applications remains challenging due to a lack of effective and high-throughput characterization techniques. Soft X-ray time-of-flight photoemission electron microscopy (XPEEM) provides element- and depth-sensitive information of materials and buried interfaces. However, chromatic and spherical aberrations cannot be corrected with electron-lens combinations. These aberrations, combined with astigmatism and space-charge effects, significantly degrade the spatial and energy resolutions. To overcome this limitation, we outline a spatial-attention based deep learning approach to automatically correct for these effects and attain nanometer resolution over the entire field-of-view (FoV). The combination of this corrective algorithm with XPEEM, termed as nanoXPEEM, establishes a new record of 48-nm spatial resolution with 232-$\mu m$ diameter FoV in the soft x-ray regime (700-1000 eV). nanoXPEEM provides unique spatial mapping of the element-specificity, depth-sensitivity, and local structure on the nanoscale. It can bridge the current gap to achieve angstrom (atomic) scale resolution.

\end{abstract}

\section{Introduction}

Two-dimensional (2D) materials have become ubiquitous for a variety of applications, ranging from classical devices to quantum information science (QIS). Their versatility enables ease of assembly and control for many applications in next-generation devices \cite{Geim13}. However, understanding and manipulating these materials effectively for real-world applications remains a challenge because of the lack of effective characterization techniques\cite{Nat24, Novoselov05}. For instance, stacking of 2D materials can scale up the throughput of transistors to many orders of magnitudes, and thus enable a new generation of electronics\cite{Montblanch23}. Precise understanding of these microscopic interactions of atoms, electrons, orbitals, and spins can give insight into material properties and effectively guide the synthesis of useful materials for devices. The time-of-flight momentum microscope (MM), when combined with ultrafast lasers and x-rays, enables both temporal- and angle-resolved microscopy, spectroscopy, and diffraction to elucidate the electron and lattice dynamics on their native length and timescales for a wide range of matter ---such as materials, nanomaterials, atoms and molecules, and warm-dense matter.


\begin{figure}[h!]
\centering
\includegraphics[width=1\textwidth]{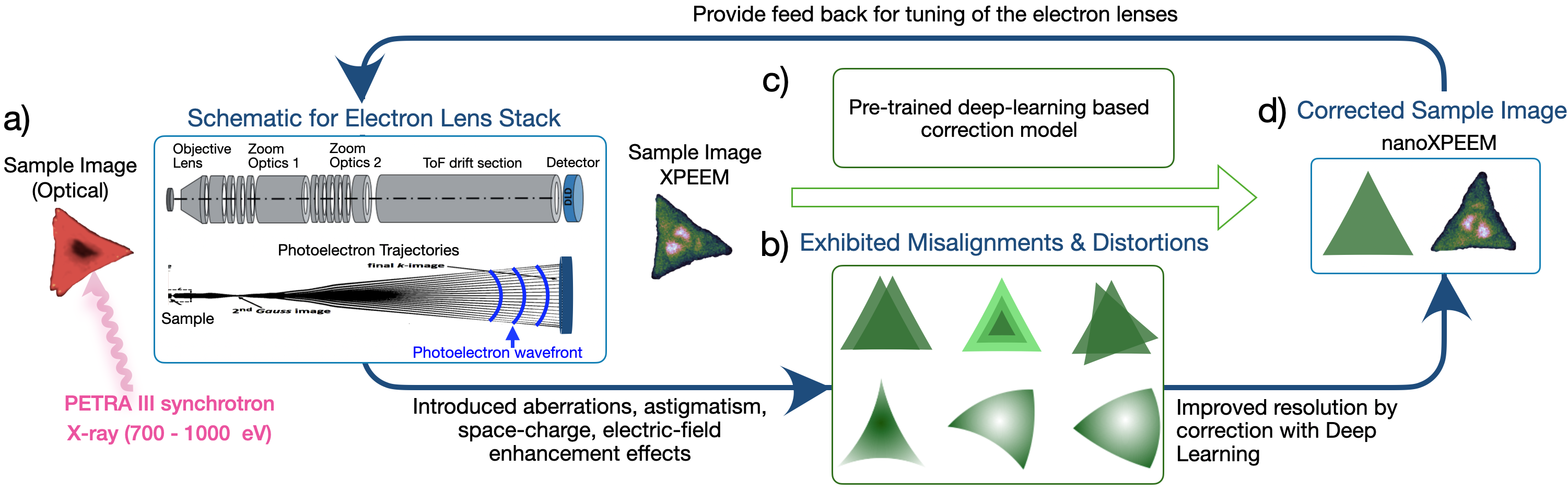}
\caption[]{\small Schematic for end-to-end correction of nonlinear effects in x-ray photoemission microscopy (XPEEM) experiments. (a) Starting from the soft-x-ray to sample interaction, where an optical image of 12\%-doped Vanadium-WS$_{2}$ is shown. We apply voltages to the electron lens stack to guide the emitted photoelectrons through the lens stack to the detector. A schematic for the lens stack with corresponding electron trajectories is shown with more details explained in Ref \cite{MetNM17, TkachAX24}. At the detector, we collect XPEEM images of W-atoms within a V-WS$_2$ flake at the W$4f$ core-levels. (b) Different types of variation and distortion introduced from the round lenses to the XPEEM images, which serve as the main limiting factors for the resolution. Specifically, time-dependent modulation, magnification difference, translation/rotation, and deformation. Similar outcomes to the XPEEM image may stem from different phenomena, such as chromatic (CA) and spherical (SA) aberrations, astigmatism, electric-field enhancement at sample edges, and space-charge. (c) We train and then apply the pre-trained deep-learning based algorithms to acquired data sets to correct for the exhibited variation and distortion to achieve nanoscale resolution for full field-of-view. (d) The corrected nanoXPEEM image of V-WS$_{2}$ at the W$4f$ core-level (33.6 eV).}
\label{fig:fig1}
\end{figure}

Until recently, MM has been demonstrated primarily with vacuum- to extreme-ultraviolet (VUV to EUV) radiation using synchrotron and tabletop high harmonic setups at up to MHz repetition rates (up to 59 eV), and with free-electron lasers (FELs) at 6 kHz at FLASH ($<300$ eV)\cite{BeaulieuPRL20, SchulerPRL22, MadeoSci20, BeaulieuSciA21, BorisenkoNC22, OvsyannikovaJES13, BenneckeAX25, allison2025cavity}. These photon energies severely undermine its field-of-view (FoV) while the repetition rate and photon flux limit the high throughput capability of MM, which offers MHz rate detection\cite{MetNM17, Tkach25Theory}. Achieving the desired signal-to-noise ratio (SNR) remains universally challenging for these different scale setups ---ranging from large facility to tabletop--- due to different factors. For example, limited x-ray beam time allocations at x-ray facilities, smaller photon energy coverage and lower photon flux for table-top light sources, and exceedingly long alignment/optimization time due to the complex experimental geometry and electron optics. 

Coupling MM with MHz laser and x-ray sources increases the SNR by a few orders of magnitude. Extending photon energy coverage into the soft-x-ray regime enables access to deep core-levels and higher penetration depth for investigating buried and heterogeneous interfaces. For example, the peak brightness and a large soft-x-ray energy coverage at LCLS-II (1 MHz, 0.25 – 1.6 keV), EuXFEL, and MHz synchrotron facilities --- PETRA III, Diamond Light Source, SSRL --- provide access to higher throughput, element- and depth-specificity \cite{TkachFragosNguyen24, SchmittDL24}. In particular, we demonstrated the coupling of MM with soft x-ray at PETRA III synchrotron radiation (0.25 - 1650 eV, 5 MHz), to map the momentum space of up to 25 Brillouin zones (up to 14 Angstrom$^{-1}$) in metals\cite{TkachU23, TkachFragosNguyen24}, time-reversal symmetry breaking of antiferromagnet \cite{FedchenkoSA24}, and chirality of kagome superconductor \cite{ElmersPRL25}, and characterization of defects in 2D materials \cite{TkachFragosNguyen24, Simoni25}. 




In this work, we demonstrate a new capability, nanoXPEEM, which pushes beyond the current horizon of PEEM to a new regime with soft-x-ray photons while achieving nanoscale resolution. To date, PEEM with $nm$-resolution has only been demonstrated with tabletop light sources that range from visible to VUV, achieving 80-$nm$ resolution in the visible (3.2 eV) \cite{LohNL18} and 20-$nm$ in the VUV (6.2 eV) \cite{StranksN20}. A complementary technique that also utilizes photoemission imaging --- nanoARPES --- demonstrated 150-$nm$ resolution by focusing the EUV light (98.5 eV) into a nano-focusing spot using a zone plate \cite{KastlAN19}. Here, we uniquely exploited the high-through advantage of MM combined with PETRA III synchrotron (0.25 - 1.6 keV, 5 MHz) to expand PEEM to the soft-x-ray regime (XPEEM). This unprecedented capability enables element- and depth-sensitive, spatial mapping of matter ---such as defects, structured systems, and interfaces, on the nanoscales.

A number of experimental and instrumental factors can severely limit the spatial and energy resolutions and the field-of-view (FoV). These include (1) electric-field enhancement at sharp sample edges, (2) space-charge effects, (3) astigmatism, and (4) time-dependent aberrations. (1) The enhancement of electric-fields at sample edges that stem from the voltage applied to the sample bias \cite{TkachAX24} and light-sample interaction. The enhancement shifts the photoelectron spectrum (XPS), which depends on the number of layers in a stacked-system \cite{Li2015, LiNano12, Fukai93}, contributing to space-charge effects. (2) Space-charge effects are impossible to overcome solely by electron optics. (3) While astigmatism can be compensated with magnetic or electric stigmators \cite{Scherzer36, schoenheJST02}, aberrations cannot be corrected by lens combinations. Applying a magnetic field to steer the photoelectrons may lead to artefacts due to additional field-particle interactions on top of the phenomena being measured. The alignment of these stigmators is also challenging and requires costly hardware optimization. (4) As described in Scherzer’s theorem \cite{Scherzer36} ---under the assumption of no space-charge--- and discussed in detail for MM \cite{schoenheJST02, TkachAX24}, the coefficients for chromatic (CA) and spherical aberration (SA) for a spherical lens stack are always positive. CA results in variable magnification in both the longitudinal (photoelectron propagation axis) and lateral (across the detector plane) directions (Figure \ref{fig:fig1}b). SA effects resemble those of photon optics, where the outer and inner electron trajectories do not focus at the same plane. All of these factors result in similar imperfections in the XPEEM images, such as misalignment (rotation and translation, magnification variation) and distortions (non-uniform deformation) (Figure \ref{fig:fig1}b). 

We performed an XPEEM experiment at the PETRA III synchrotron with soft-X-ray photons ranging from 250 - 1000 eV to characterized vanadium defects in WS$_2$ samples, where we utilized the nanoXPEEM procedure. Details on the scientific outcomes are described in our recent manuscript\cite{TkachFragosNguyen24} and an upcoming manuscript\cite{Simoni25}. In this article, we address the imperfections exhibited in the acquired XPEEM data, at ~820 eV photon, to (1) improve the spatial and energy resolutions and (2) compute correction metrics ---which will inform the alignment of electron optics in the near-future to overcome experimental hurdles. Specifically, we correct for XPEEM imperfections using deep-learning (DL) approaches. This methodology can also benefit a wide range of imaging techniques and scientific fields, such as material and nanoscience, QIS, and high energy density science.



\newpage

\section{Methods}

\begin{wrapfigure}{r}{0.6\textwidth}
  \begin{center}
    \includegraphics[width=0.6\textwidth]{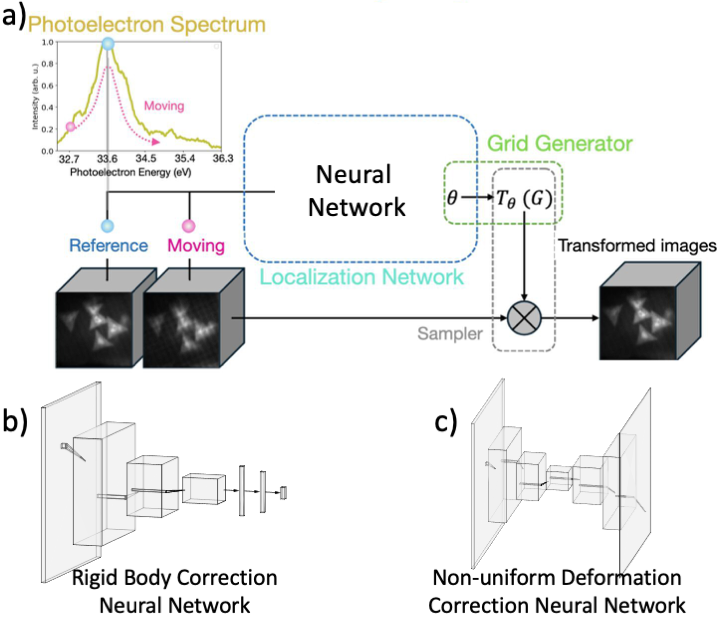}
  \end{center}
\caption[]{\small Schematics for deep-learning correction approach, including a) uniform rigid body (translation, rotation, and magnification) and b) non-uniform deformation correction. During training, a set of two random XPEEM images (reference (blue) and moving (pink)) are inputted to the localization network, which learns a parametric mapping to transform the moving image to minimize the mean squared error difference of moving and reference images. During prediction, the XPEEM image at the peak of the XPS acts as the fixed $I_{reference}$, while all remaining XPEEM images along the x-ray photoelectron spectrum ($I_{moving}$) are aligned to this reference. The generator transforms a structured grid using the learned mapping. The sampler uses bilinear interpolation to sample the intensities of a moving image at the transformed pixel locations, which results in a transformation of the moving image.}
\label{fig:fig2}
\end{wrapfigure}

Based on the simulated electron trajectories, under the assumption of no space-charge effects and aberrations, the spatial resolution at this experimental condition is sub-60 $nm$ \cite{BschoenNJP18, TkachAX24, Simoni25}. These simulations predict that we can further achieve 20-nm with this state-of-the-art MM filtering out photoelectron signals based on their energies. However, removing photoelectrons will lower SNR and require impractical acquisition times. This approach is currently not feasible due to low SNR and detrimental linear/non-linear effects --- aberrations, astigmatism, field-enhancement, and space-charge. 

Instead, we apply the Spatial Transformer Network (STN) \cite{jaderberg2015spatial} to correct for the aforementioned mechanisms that limit the resolution. The original STN and its variants were used for image alignment in bio-medical analyses \cite{li2018non, kuang2019faim, de2017end}. We extend this network to develop nanoXPEEM (Figure \ref{fig:fig2} a, b). Here, the localization neural network learns the parameters of a pre-supposed transformation to align the moving images ($I_{moving}$) to the reference image ($I_{reference}$). The XPEEM image at the peak of the XPS acts as a fixed $I_{reference}$, while all remaining XPEEM images along the XPS ($I_{moving}$) are aligned to this reference. 

We perform the correction sequentially with (1) rigid body correction for the rotation, translation, and magnification variation, and (2) non-uniform deformation correction for distortions. 



\noindent \textbf{Rigid body (Rotation and Translation) Correction}

The rigid body correction applies an affine transformation to the moving images using the parameters predicted by an STN. This alignment algorithm acts on a set of image pairs that consists of $I_{reference}$ and $I_{moving}$ (Figure \ref{fig:fig2}a). The localization network takes in tuples of images, $\mathbb{R}^{2 \times H \times W}$, and consists of 4 convolution blocks (each consisting of a convolution layer, non-linearity and max-pooling), followed by two dense layers. This network learns the six parameters of the affine transform, $\theta$, to be applied to $I_{moving}$, defined as


\begin{equation}
\begin{bmatrix}
x^{s}_{i} \\
y^{s}_{i} \\
\end{bmatrix}
         = \begin{bmatrix}
         \theta_{11}~~~\theta_{12}~~~\theta_{13} \\
         \theta_{21}~~~\theta_{22}~~~\theta_{23} \\ 
         \end{bmatrix}
         \begin{bmatrix}
x^{t}_{i} \\
y^{t}_{i} \\
1
\end{bmatrix},
\end{equation}
$(x^{s}_{i}, y^{s}_{i})$ are the source coordinates in the input feature map that defines the sample points, and $(x^{t}_{i}, y^{t}_{i})$ are the target coordinates in the output. The generator defines a regular grid in the output feature map, and transforms it according to the learned affine transformation. The bilinear interpolation sampler interpolates the pixels of the original $I_{moving}$ to the locations of the transformed grid, which then generates a transformed $I_{moving}$. The loss function (Eq 2) is the sum of squared intensity differences between the pixels of $I_{reference}$ and the transformed $I_{moving}$. Here, $Q_{\theta}$ is the affine transformation parametrized by $\theta$ and $||.||^2$ represents the sum of squares over all image pixels. 

\begin{equation}
L(I_{reference}, I_{moving}) = || I_{reference} - Q_{\theta}(I_{moving}) ||^2,
\end{equation}


\noindent \textbf{Non-uniform Deformation Correction} 

The non-uniform deformation correction (Figure \ref{fig:fig2}b), takes additional steps beyond rigid body to account for nonlinear deformation and distortion effects, including CA and SA, field enhancement, and space-charge. 

This correction also uses the STN, with two differences. First, the network takes the rigid body corrected images as inputs rather than the raw images. Second, while the rigid body localization network outputs a six-dimensional vector to parametrize an affine transform, this localization network is an autoencoder that outputs a deformation field of the same size as the input images, $\mathbb{R}^{H \times W}$. This deformation field is added to the pixel coordinates of $I_{moving}$ to get their transformed locations. Then, bilinear sampling determines the intensities of the pixels at their new locations. While training the network, we use the mean squared error loss as before, along with a regularization parameter that penalizes the norm of the deformation field. 

Both NNs are trained to minimize differences between pairs of images. The intention is to adjust the moving image to remove deformations due to aberration, but because the images correspond to different layers of the sample, it is possible that the NNs will inadvertently remove true physical differences between the layers. We apply regularization to reduce the risk of over-correction, and post-analysis (see section below) shows that the deformation fields are consistent with known space-charge aberrations. In the supplemental section, we provide a heuristic validation of the non-uniform aberration correction algorithm.

Prior researchers have investigated data driven Machine Learning based approaches to improve the image quality from XPEEM/LEEM. A focus of this effort has been on unsupervised denoising the data using for instance Singular Value Decomposition (SVD) \cite{masia2013quantitative, masia2023low}, Bayesian Unmixing \cite{nemvsak2018aqua}, Variational Autoencoders \cite{restrepo2022denoising}, etc. The objective of our investigation is not denoising, but aberration correction. Researchers have attempted to ameliorate the impact of such aberrations in XPEEM using dynamic correcting lenses \cite{schoenheJST02}, sample coatings that act as energy filters \cite{ishiwata2012diamondoid}, the use of electron mirrors \cite{tromp2013new}, etc. We attempt this aberration correction using self-supervised deep learning. This acts as a post-processing stage and does not require any changes to the experimental apparatus or the setup.



\noindent \section{Discussion \& Results}
 
Once the photoelectrons are emitted from the sample surface post light-matter interaction, the electron optics guide them to a circular delayed-line detector (Figure \ref{fig:fig1}a) \cite{MetNM17}. Hardware filters unwanted photoelectrons by their kinetic energies \cite{TkachJS24, TkachAX24}. In this configuration, the magnification reaches 120x at 824 eV photon energy to enable sub-60 nm resolution under perfect condition (Figure \ref{fig:fig1}b). However, our measurements \cite{Simoni25} yield 250-nm resolution due to the effects described above ---astigmatism, field enhancement, aberrations, and space-charge. 

\begin{figure}[h!t]
\centering
\includegraphics[width=1\textwidth]{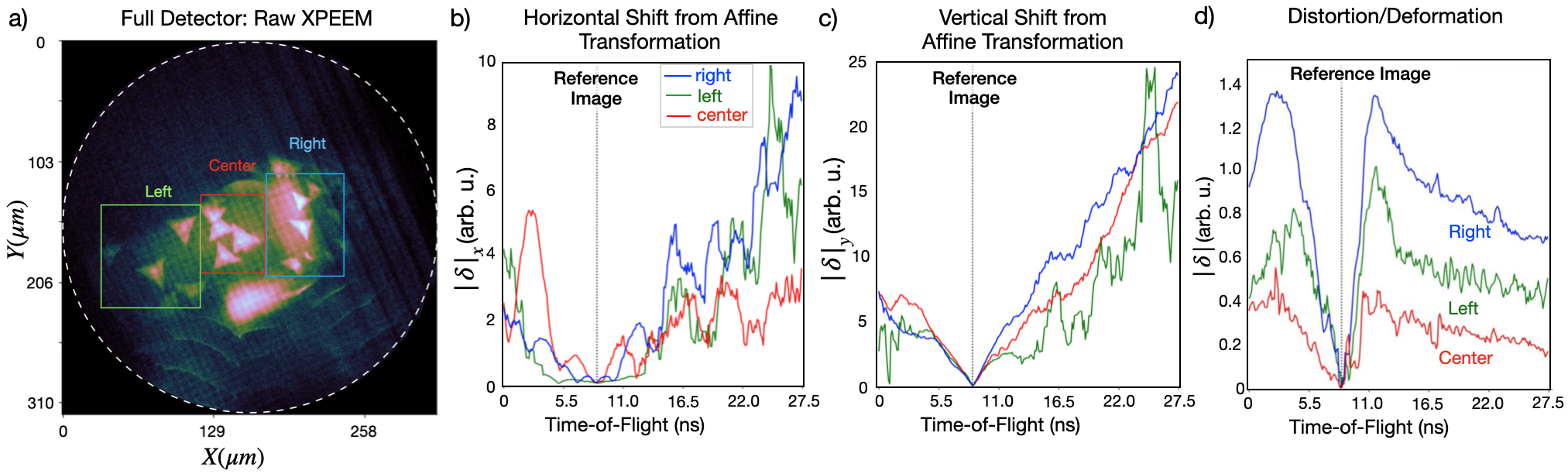}
\caption[]{\small
(a) Acquired XPEEM image of the full detector for W$4f_{5/2}$ core-level, which sums over the entire stack of images acquired over 33 ns time-of-flight range. We divided it into three regions: Left (green), Center (red), Right (blue), to apply the DL-corrections separately as they experienced different linear and nonlinear effects that cause rotation, translation, magnification variation, and distortions to the XPEEM images. The resulting asymmetry in the photoelectron wavefront result in significant distortion around the center of the detector, particularly the Left and Right areas. (b) Horizontal and (c) vertical (on the detector plane) corrective shift magnitude to the XPEEM images, which are extracted from the Rigid Body Correction for all three regions. (d) The norm ($l_2$) of the deformation field for the XPEEM images in the corresponding regions that represent the non-uniform Deformation Correction magnitude. In panels (b-d), the vertical dotted line (grey) indicates the reference image for the Rigid Body and non-uniform Deformation Corrections as described in Methods. The reference image corresponds to the focal point of the electron lenses, which is also the XPEEM image at the peak intensity of the W$4f_{5/2}$ XPS. 
}
\label{fig:fig3}
\end{figure}

Due to these underlying mechanisms that significantly degrade the resolution, as discussed in the introduction and Figure \ref{fig:fig1}b, Figure \ref{fig:fig3} a-d show the quantitative analysis of these effects on the measured images of W$4f_{5/2}$ core-level. Figure 3a displays the full XPEEM detector image as a sum of images collected over 33-ns time range. In perfect condition with no linear and non-linear particle interaction effects, the wavefront of photoelectrons is symmetric when arriving at the detector. However, these effects distort the photoelectron wave-fronts and cause asymmetries. These asymmetries leads to non-uniform distortions in the image that requires a set of different parameters for training the neural net for deformation correction. Hence, we divide each XPEEM image into three areas to apply corrections separately, namely: Left, Center, Right.

We observe stronger distortion in the photoelectron wavefront at regions away from the center of the detector (Figure 3b-d). In particular, the Right and Left regions yield a higher order of translation/rotation, magnification variation, and deformation compared to the Center region. Each point on the curves corresponds to the norm of an XPEEM image at each TOF slice. As originally discussed in Ref \cite{schoenheJST02}, the longitudinal and lateral CA lead to a time-dependent breathing mode in the XPS, which we quantitatively extract from the rigid body correction (Figures \ref{fig:fig3} b,c) for the three regions on the detector. CA also causes variation in the magnification that results in a variation on the real-image plane for different photoelectron energies, which likely gives rise to the rising slope in the panels (b) and (c). Here, the reference image corresponds to the focal point of the electron lenses, which is also the XPEEM image at the peak intensity of the W$4f_{5/2}$ XPS. 

\begin{figure}[h!t]
\centering
\includegraphics[width=0.63\textwidth]{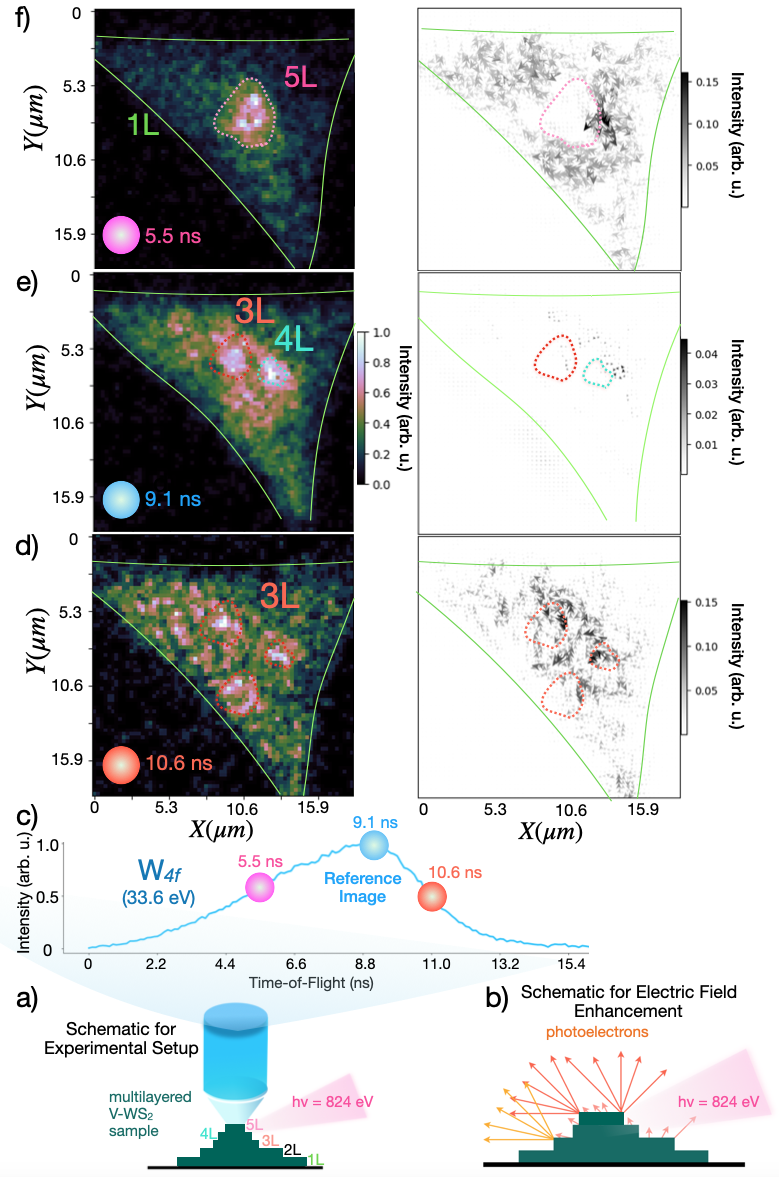}
\caption[]{\footnotesize
(a) Schematic of XPEEM setup with multilayered V-WS$_2$ sample and 824 eV x-ray photons.(b) Schematic diagram for electric-field enhancement from the edges of the multilayered system likely resulting from the strong field applied at the sample position (1.3 kV/mm to 3.8 kV/mm) \cite{Simoni25, TkachU23, TkachAX24}. Similar enhancement effect was observed in for multilayered TMDs that result in an energy shift of the corresponding photoelectron spectrum (XPS) \cite{Li2015, LiNano12, Fukai93}. (c) XPS of W$4f_{5/2}$ core-level as a function of electron time of flight. (e)The XPEEM image at the peak (blue circle) of the spectrum is used as a shape-reference for the correction algorithms. (d, f) An example of moving XPEEM images (pink \& red circles)  that are corrected based on the reference. The quiver plots show the direction \& magnitude of the correction, which indicate that most corrections occur near the mulitlayered islands, where the electric-field enhancement occurs as depicted in panel (b). In (d-f) green lines are guide to eye to show the monolayer of (1L) of V-WS$_2$. Dotted lines are added to show the multilayer islands of three (3L), four (4L), and five (5L) stacked layers. The same guides to the eye for mono- and multilayer stacks are also shown on the corresponding deformation map, which confirm that most deformation corrections take place at the edges of the multilayer islands to compensate for field-enhancement and accompanied space-charge effects. 
}
\label{fig:fig4}
\end{figure}

Meanwhile, SA, electric-field surface enhancement, and space-charge effects lead to non-uniform distortion/deformation and rotation in the XPEEM images (Figure \ref{fig:fig1}d). These detrimental outcomes collectively become more significant for $I_{moving}$ as they approach $I_{reference}$ (Figure \ref{fig:fig3}d). Figure \ref{fig:fig4} depicts the schematic and quantitative analysis of this phenomenon, where panels (a) and (b) show the schematics of the experimental setup and electric-field enhancement that takes place at the sharp edges of the multilayer islands \cite{Li2015, LiNano12, Fukai93}. We expect that these hot spots contribute to the space-charge and lead to non-uniform distortion of XPEEM images (Figures \ref{fig:fig3}d \& \ref{fig:fig4}b), which is then compensated by the non-uniform deformation correction (Figure \ref{fig:fig4}). 

The resulted W$4f_{5/2}$ XPS (Figure \ref{fig:fig4}c) with corresponding the 2D deformation maps show the correction taking place at different TOF slices --- at the peak and full-width at half maximum (5.5, 9.1, and 10.6 ns) (Figure \ref{fig:fig4}d-f). They consist of arrows representing the directionality and magnitude of the deformation correction. The primary corrections take place around the edges of the multilayer islands, which are located in the central region of the flake, with the monolayer corresponding to the large triangular shape (Figure \ref{fig:fig4}d-f ). These high charge density regions govern the topology of the deformation field. Specifically, the deformation field magnitude and direction correspond to the integral and iso-contour of the charge density, respectively. Additionally, the distance between  $I_{moving}$ to $I_{reference}$ dictates the average magnitude of the deformation field. Figure \ref{fig:fig4}e shows the input  $I_{reference}$ and the corresponding deformation map with negligible correction, which is due to the fact that the DL-algorithm is self-referenced.  

\noindent \subsection{Quantitative Comparison for DL-corrections versus measurements} 

As the DL-corrections are self-contained/referenced, we benchmark their performance using unsupervised quality metrics and line-outs with error function fits, as detailed below. 

\noindent \textbf{Total Variation  Score (TV).} The TV score 
\cite{mittal2012no} estimates the complexity of an image with respect to its spatial variation,  given by
$TV(X)=\sum_{i,j\in\mathcal{N}} \|x_i-x_j\|_2$,
where $\mathcal{N}$ defines the neighborhood of the pixel and $\|.\|_2$ denotes the $l_2$ norm. Images with lower quality ---low-contrast \cite{estrela2016total} and noisy--- have higher TV scores. By applying the DL-corrections, we achieve better alignment of all images in the stack. The edges of each sample flake at different TOF become better aligned and yield much higher contrast. A byproduct of these corrections includes washing out of the noise pixels, which also reduces the TV scores. 

\noindent \textbf{Difference of differences of the median-filtered image ($\Delta$DoM).} 
$\Delta$DoM \cite{kumar2012sharpness} estimates the sharpness/blurriness of a given image by examining the contrast at the edges of the sample. A well aligned image stack will have sharper edges, and a higher $\Delta$DoM sharpness score.

\noindent\textbf{Line-outs and error function fit}. We extract a line-out at the same location from an edge of a V-WS$_2$ flake (Figure 5c) from the XPEEM and corresponding nanoXPEEM images. Then, we apply an error function fit to determine the spatial resolution, which correspond to the 20-80\% width (Figure 5d, Table 1). 


\begin{figure}[h!t]
\centering
    \includegraphics[width=1.05\textwidth]{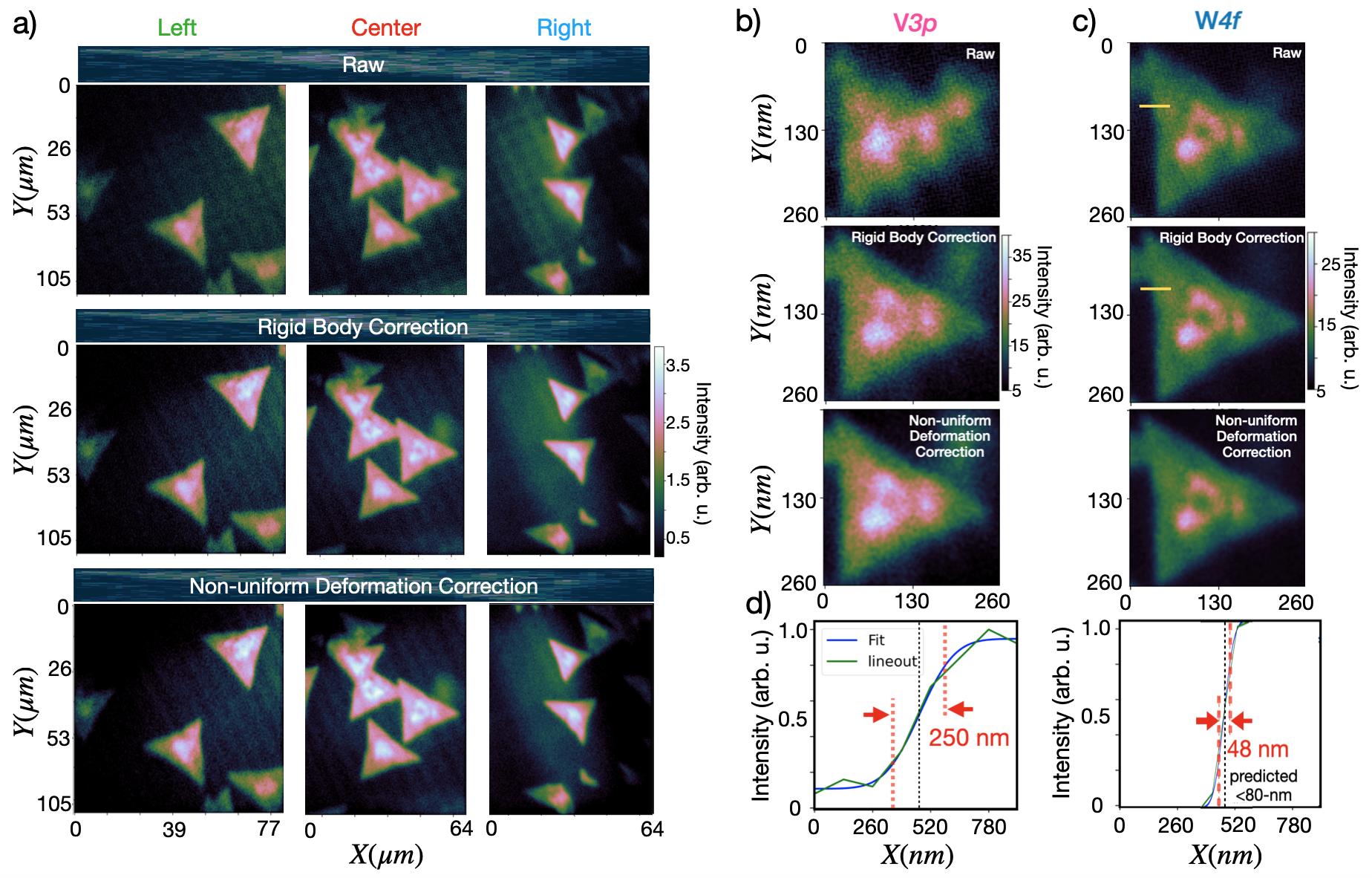}
\caption[]{\small  Comparison of raw (XPEEM) and corrected (nanoXPEEM) images: (a) Raw, Rigid Body, and Non-uniform Deformation corrected images of V-WS$_2$ flake for the Left (green), Center (red), Right (blue) regions at the W$4f_{5/2}$ core-level. A set of zoom-in images of a single flake in the center region, acquired at the (b) V$3p$ (37.2 eV) and (c) W$4f_{5/2}$ (33.6 eV) core-levels. (d) Line-outs with fits applied to the XPEEM and nanoXPEEM images, where the yellow line in panel (c) indicates the location of the line-outs. The instrument resolution is 250-nm, which improves by 5x with the Rigid Body and Non-uniform Deformation corrections to achieve a record 48-nm resolution.
}
\label{fig:fig5}
\end{figure}

All of these metrics --- TV, $\Delta$ DoM, and line-outs --- yield excellent agreement that showcase strong improvements for contrast, resolution, and FoV offered by the DL-corrections compared to raw images (Table \ref{table:1}). Both DL-corrections combined remove the severe limitation in the resolution across the detector, especially in the areas around the Center region (Figures \ref{fig:fig3} \& \ref{fig:fig5}). The resolution of XPEEM ranges from 250-nm to 2.3-$\mu m$ (Table \ref{table:1}), depending on the detector region and SNR level. At extremely low SNR combined with severe distorted photoelectron wavefront (e.g. for V$3p$ in the Left region), the resolution is degraded to 2.3-$\mu m$. By applying both DL-corrections, we achieve nanometer resolution over the entire detector. Specifically, we achieve around or sub-100-$nm$ across a wide-FoV of 232-$\mu m$, with a record of 48-$nm$. This is a significant improvement from XPEEM where only the Center region (132-$\mu m$ FoV) reached nanoscale, with the best record of 250$-nm$ resolution. 

Overall, the DL-corrections combined improve up to 20x from XPEEM to nanoXPEEM. Figure 5b-c substantiates the robustness of these DL-corrections for handling low SNR data. Specifically, the peak XPS intensity of W$4f_{5/2}$ is 30x stronger than for V$3p$ \cite{TkachFragosNguyen24, Simoni25} due to the difference in cross-section. By applying the DL-corrections, we still attain substantial improvement for V$3p$, from 1.5 $\mu m$ to 77.4 $nm$ resolution in the Center region. Meanwhile, the XPEEM acquired at W$4f_{5/2}$ yields a 250-nm resolution, which is improved to 48-nm after correction. Comparison of XPEEM with nanoXPEEM images yield a distinct contrast in the SNR, where the SNR increases by 17\% for post-rigid body and 42\% for post-deformation correction (Figure S3 in SM and  \ref{fig:fig5}a-c). 

The trained-DL models and acquired knowledge from these corrections, particularly the photoelectron footprints on the detector (Figures \ref{fig:fig3} \& \ref{fig:fig4}), provide unique insights for future optimization of complicated material systems and live-tuning of the instrument. For example, we can use the pre-trained DL-models to predict the voltage settings of the electron optics for multilayered and structured devices. The pre-trained models can also enable efficient and effective live-tuning of the instrument to optimize for nanoXPEEM images during an experiment (Fig \ref{fig:fig1}). Moreover, the significant improvement in SNR reduces the acquisition time for acquiring data with similar or much better signal level within less or similar time range. These improvements enable element- and depth-selective imaging that were previously impossible due to the low resolution, SNR, and limited time at x-ray facilities. 

\begin{table}
\footnotesize
  \centering
  \setlength{\tabcolsep}{12pt}
  \begin{tabular}{llccc}
 \hline
    & \textbf{Metric} & \textbf{Raw}  & \textbf{Rigid Body}  & \textbf{Non-uniform Deformation}  \\[1pt] 
    & &               & \textbf{Correction}  & \textbf{Correction}  \\[1pt] 
  \hline     
     \textbf{W$4f$ Right}   &   Total Variation ($\downarrow$) & 1.42  & 0.567 & 0.368 \\ 
                            &   $\Delta$DoM ($\uparrow$)       & 0.640 & 0.924 & 0.967 \\ 
                            &   Resolution ($\downarrow$ nm)   & 1057  & 490   & 116 \\ 

  \hline     
     \textbf{W$4f$ Center}  & Total Variation ($\downarrow$)   & 1.258 & 0.677 & 0.421 \\ 
                            & $\Delta$DoM  ($\uparrow$)        & 0.651 & 0.912 & 0.952 \\ 
                            & Resolution ($\downarrow$ nm)     & 245   & 48.6  & 52.5 \\ 

  \hline     
     \textbf{W$4f$ Left}    & Total Variation ($\downarrow$)  & 1.498 & 0.716 & 0.511 \\ 
                            & $\Delta$DoM ($\uparrow$)        & 0.712 & 0.912 & 0.962 \\ 
                            & Resolution ($\downarrow$ nm)    & 255 & 154.8 & 83.8 \\ 

  \hline     
    \textbf{V$3p$ Right}    & Total Variation ($\downarrow$)  & 2.712 & 1.368 & 0.647 \\ 
                            & $\Delta$DoM ($\uparrow$)        & 0.714 & 0.923 & 0.941 \\ 
                            & Resolution ($\downarrow$ nm)    & 361 & 142 & 70.5 \\ 

  \hline     
     \textbf{V$3p$ Center}  & Total Variation ($\downarrow$) & 3.426 & 1.571 & 0.692 \\ 
                            &   $\Delta$DoM ($\uparrow$)     & 0.713 & 0.931 & 0.949 \\ 
                            &   Resolution ($\downarrow$ nm) & 1548  & 387   & 77.4 \\ 

  \hline     
     \textbf{V$3p$ Left}    & Total Variation ($\downarrow$) & 3.533 & 1.416 & 0.534 \\ 
                            & $\Delta$DoM ($\uparrow$)       & 0.740 & 0.997 & 1.046 \\ 
                            & Resolution ($\downarrow$ nm)   & 2322  & 258   & 168 \\ 

  \hline     
     \textbf{Au$4f$ Right}  & Total Variation ($\downarrow$) & 1.558 & 0.414 & 0.372 \\ 
                            & $\Delta$DoM ($\uparrow$)       & 0.616 & 0.664 & 0.774 \\ 
                            & Resolution ($\downarrow$ nm)   & 503 & 77.4 & 77.4 \\ 

  \hline     
     \textbf{Au$4f$ Left}   & Total Variatio ($\downarrow$) & 1.237 & 0.572 & 0.477 \\ 
                            & $\Delta$DoM ($\uparrow$)      & 0.623 & 0.649 & 0.711 \\
                            & Resolution ($\downarrow$ nm)  & 490.2 & 167.7 & 64.5 \\ 
  \hline		 
  \end{tabular}
  
  \caption{Comparison of metrics  --- TV scores, $\Delta$ DoM, and line-outs --- for different raw and corrected V-WS$_2$ flakes using Rigid Body and Non-uniform Deformations. As indicated by the symbols, for $\Delta$ DoM, higher values are better. For Total Variation and resolution, lower values are better. All the metrics are in excellent agreement regarding the improvement of the corrections compared to before.\label{table:1}}
\end{table}


\section{Conclusions}

2D materials have been used for numerous electronic and device applications. However, applying and scaling these systems for industrial applications remain challenging due to lack of effective characterization techniques. We recently demonstrated XPEEM to offer high-resolution and element-sensitivity measurements for 2D materials \cite{TkachFragosNguyen24, Simoni25}. In this work, we take a further step to combine XPEEM with deep-learning to develop nanoXPEEM for the soft-x-ray regime, which provides unique microscopy capability for element-specificity, depth-sensitivity, and local structure mapping on the nano-scale. nanoXPEEM can bridge the gap from the nanometer to angstrom (atomic) scale mapping by significantly improving the throughput and resolution. The developed approach can also apply to the momentum imaging mode of this instrument to attain high-throughput band structure tomography with higher energy resolution. Additionally, the pre-trained DL-models can enable efficient live-tuning of the instrument to enable optimization of signals for complex materials and devices. This approach is broadly applicable to a wide range of x-ray and electron microscopy techniques for a large range of matter.




\begin{acknowledgement}

We thank Chi-Chang Kao for reviewing the manuscript. A.M. and Q.L.N were supported by the Department of Energy, Laboratory Directed Research and Development program at SLAC National Accelerator Laboratory, under contract DE-AC02-76SF00515. Q.L.N. thanks the Bloch Fellowship in Quantum Science and Engineering by the Stanford-SLAC Quantum Fundamentals, Architectures and Machines Initiative. 

\end{acknowledgement}




\bibliography{achemso-demo}

\end{document}


\newpage

\section{Non-uniform Deformation Correction}
Deep learning models are trained with random initializations of the model parameters, dataset order, batching of data samples, etc. In this vein, the quality of the minima that the learning algorithm converges to (and thus the reproducibility of their predictions) may be problematic. Additionally, data driven models are trained to minimize loss over data and thus, can lead to greedy behavior that focuses on minimizing loss instead of conforming to physics. In this context, correct regularization in the spatial transformer model loss is important, especially for the non-uniform deformation correction. Here, we present supplementary results to address these issues. 

\begin{figure}[h!t]
\centering
\includegraphics[width=1\textwidth]{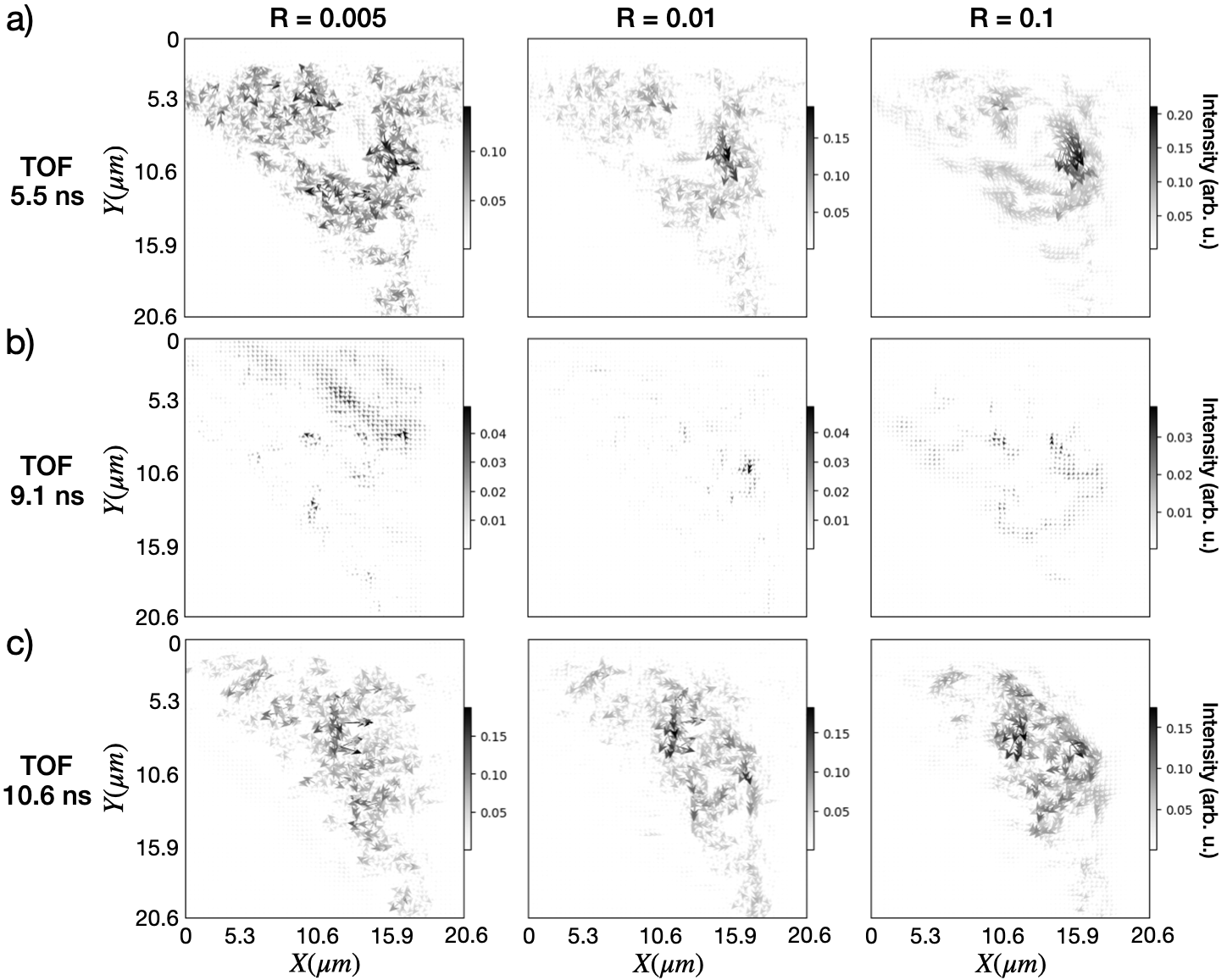}
\caption[]{Comparison of the non-uniform deformation fields predicted by the methodology at varying values of the regularization parameter
(a) comparison for TOF of $5$ ns,
(b) comparison for TOF of $8.3$ ns,
(c) comparison for TOF of $9.6$ ns.
}
\label{fig:figS1}
\end{figure}

In Figure \ref{fig:figS1}, we show the non-uniform deformation correction with different degrees of regularization, explicitly the value of the regularization parameter, reported in the figures as $R$. This is a hyper-parameter of the learning algorithm and its value is determined via validation. Here, we report the deformation fields while varying the parameter by two orders of magnitude. We observed similar behavior from the correction to compensate the surface enhancement effect caused by the high voltage applied to the extractor (1-3 kV). $R=0.01$ is chosen as the correction reported in Figure 4 of the main text. With a increasingly higher values of the regularization parameter, we expect the the deformation field being forced to be smoother and sparser due to penalization, as is observed in the figure. The final value of the regularization parameter was selected at $R=0.01$ based on cross-validation and domain knowledge.

\begin{figure}[h!t]
\centering
\includegraphics[width=1\textwidth]{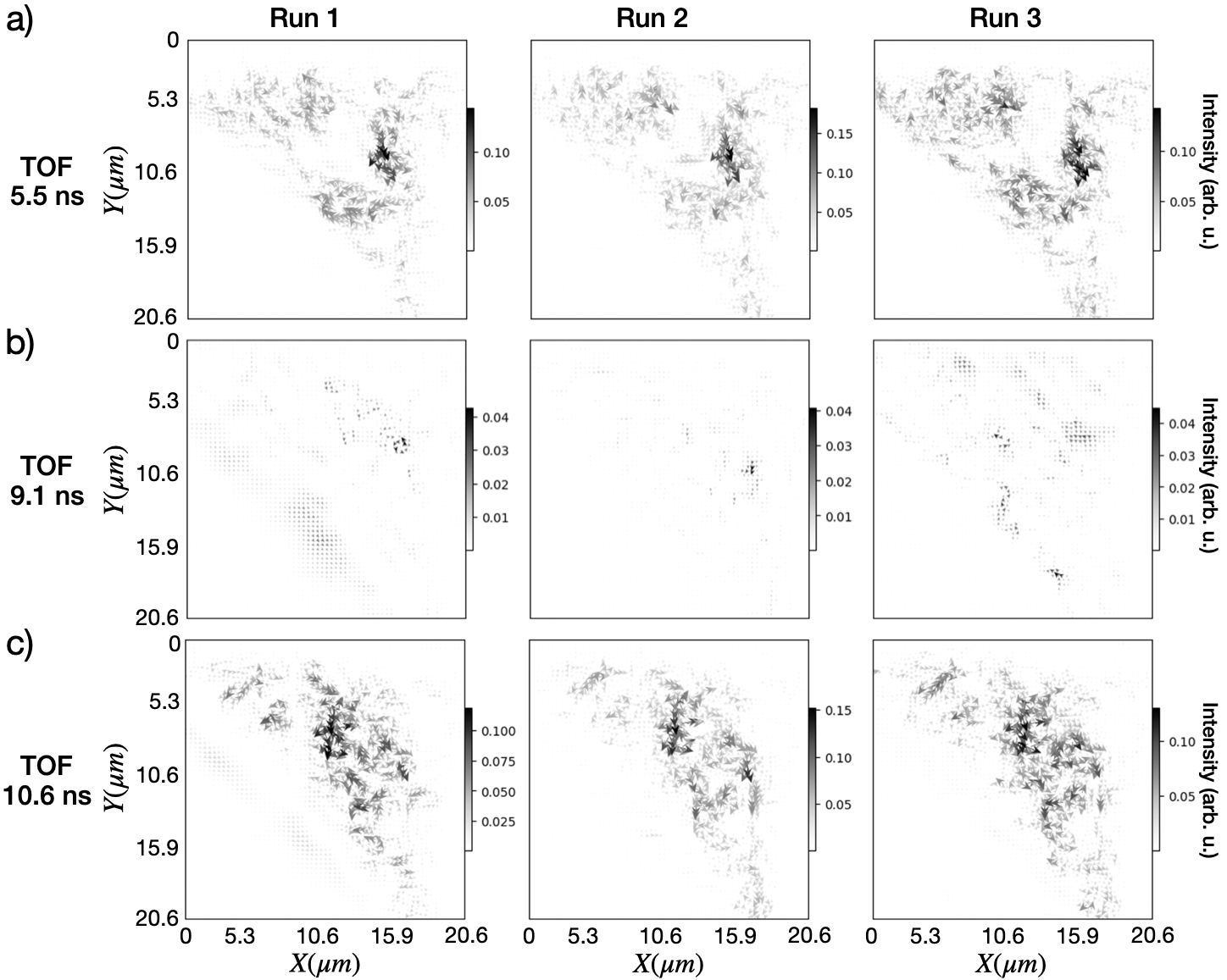}
\caption[]{Comparison of the non-uniform deformation fields predicted by the methodology at $R=0.01$ across multiple runs with different initializations
(a) comparison for TOF of $5$ ns,
(b) comparison for TOF of $8.3$ ns,
(c) comparison for TOF of $9.6$ ns.
}
\label{fig:figS2}
\end{figure}

To confirm the reproducibility of the non-uniform deformation alignment methodology, we repeated the same calculation at least three times (Figure \ref{fig:figS2}). This tests for the stability and robustness of the predicted deformation fields, and thus also the final aligned XPEEM slices.  We observe in Figure \ref{fig:figS2} that all the different runs yield very similar deformation fields, both qualitatively and quantitatively. We performed the same repeats for different TOFs at 5, 8.3, and 9.6 ns.

The difference between the peak slice and any other slice arises due to two sources: the differing levels of aberration between the two slices that is unphysical, and other differences that arise due to physics. Bereft of any inductive bias that can enable the neural network to differentiate between these two differences that are diametric in nature, there is a possibility that the non-uniform, spherical aberration correction procedure would correct for both kinds of differences, resulting in corrected slices that may not adhere to underlying physics. In this scenario, the algorithm would resort to a greedy behavior where it would deform the pixels with the highest intensity, residing in the interior of the multilayer islands with multiple stacked layers. In Figure 4 of the main text, and in Figures 1 and 2 of the supplemental section, we observe that within the range of regularizations used in this investigation, such deformations are penalized adequately and the algorithm does not deform the interior of these multilayer islands at all, but focuses on the edges of the multilayer islands where the spherical aberration effects are important. In physics terms, the corrections from the spherical aberration correction algorithm are focused at the edges of the multilayer islands so as to compensate for field-enhancement
and accompanied space-charge effects. This serves as a heuristic validation of the spherical aberration correction algorithm used in this investigation, while also highlighting the need for judicious choice of the regularization for this algorithm. 

\begin{figure}[h!t]
\centering
\includegraphics[trim={0cm 3cm 0cm 1.5cm},clip,width=1\textwidth]{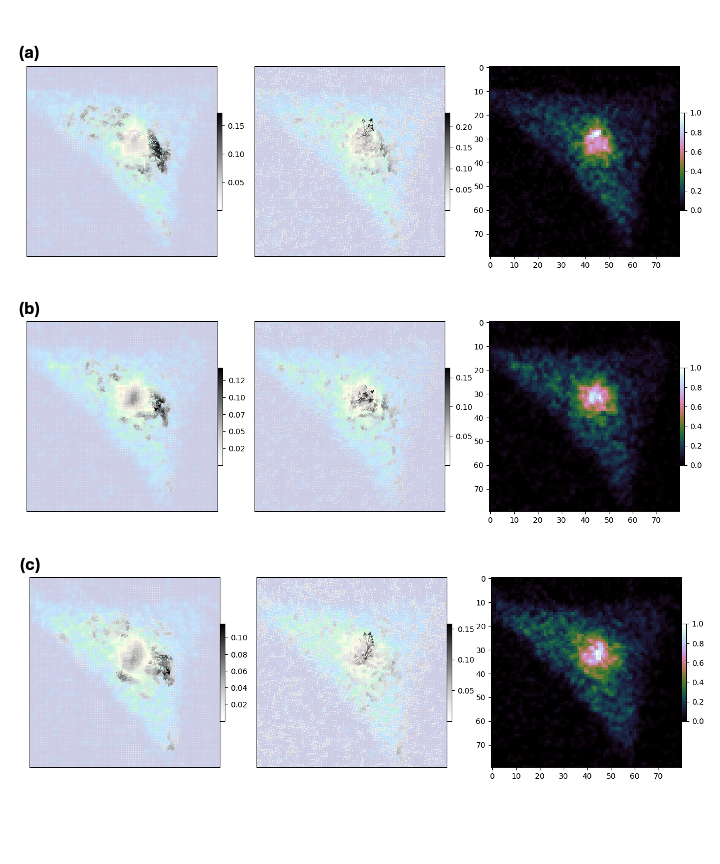}
\caption[]{Comparison of the non-uniform deformation maps predicted by the methodology for $R=0.0$ against the case with $R=0.01$ for given input XPEEM slice regions. For each figure, from left to right, we exhibit the case with $R=0.01$, the case with $R=0.0$ at the center and the input XPEEM slice region on the right.
(a) comparison for TOF of $5.5$ ns,
(b) comparison for TOF of $5.9$ ns,
(c) comparison for TOF of $6.5$ ns.
As is seen, the unregularized algorithm focuses on deformations in the interior of the multilayer islands, while the regularized algorithm focuses on the edges of the multilayer islands.
}
\label{fig:figS22}
\end{figure}

\section{Details of a two-step procedure for correction}
While the deformation field approach used in the non-uniform correction step could also learn to account for the rigid body alignment, we observed inconsistent performance using such a single-shot approach due to the small datasets and the inherent noise in the data. Using a two-step approach focusing solely on the rigid body alignment first, and then correcting for non-uniform deformations yielded better results on all metrics consistently. Consequently, this two-step approach is used in the investigation and the results reported.

\newpage

\section{Comparison of signal to noise (SNR)}
\begin{figure}[h!t]
\centering
\includegraphics[width=0.5\textwidth]{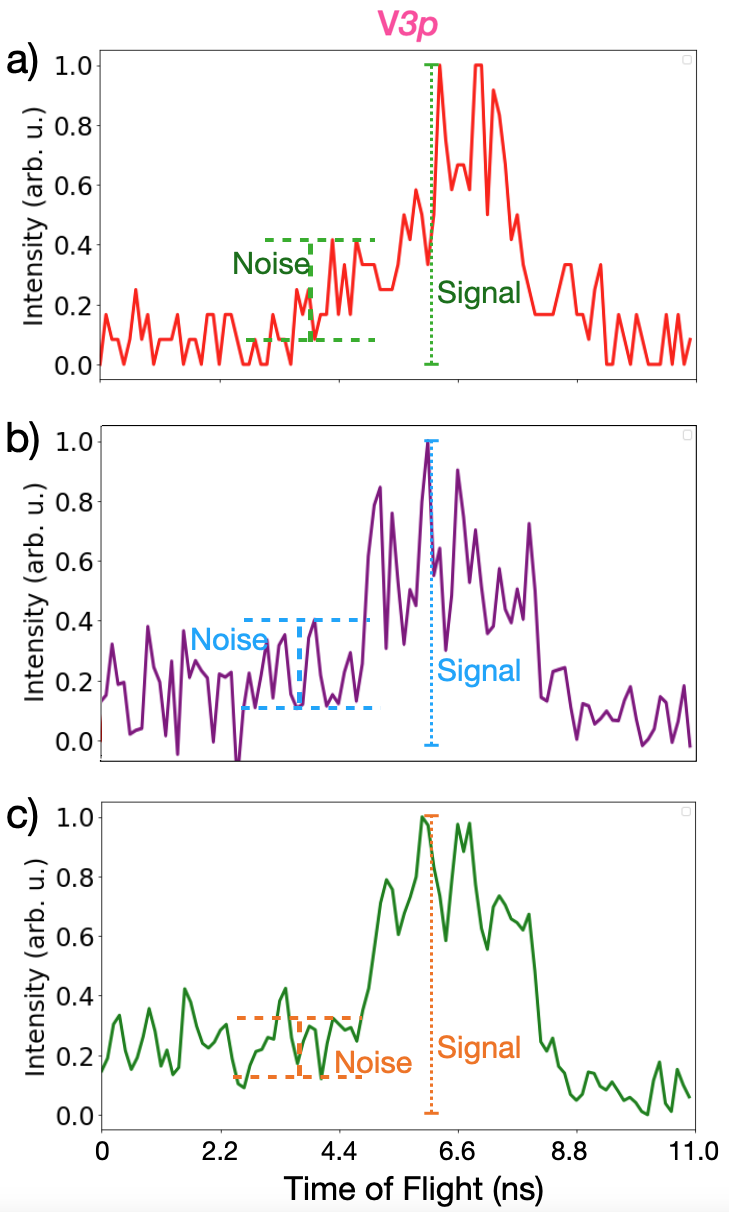}
\caption[]{
Extracted line-out at the same position for the (a) raw XPEEM images, (b) images with Rigid Body correction, and (c) images with both Rigid Body and non-uniform deformation correction. The signal and noise levels are labeled in each panel with dotted lines, where the signal is normalized to one. The noise level clearly decreases after each correction layer that results in 17\% and 42\% improvement in the SNR after we applied the Rigid Body and non-uniform deformation corrections. 
}
\label{fig:figS3}
\end{figure}

The SNR level between the raw data with corrected data using both Rigid Body and non-uniform Deformation corrections are shown in Figure \ref{fig:figS3}. Here, we define $SNR = (S-N)/N$, where the noise and signal levels are labeled in \ref{fig:figS3}.

